# "Cooking carbon in a solid salt": Synthesis of porous heteroatom-doped carbon foams for enhanced organic pollutant degradation under visible light

Jiang Gong [a], Jinshui Zhang [b], Huijuan Lin [c], Jiayin Yuan [d,e],*

[a] Department of Chemistry, University of Texas at San Antonio, One UTSA Circle, San Antonio, Texas 78249, USA
[b] State Key Laboratory of Photocatalysis on Energy and Environment, College of Chemistry, Fuzhou University, Fuzhou 350116, P. R. China
[c] Key Laboratory of Flexible Electronics (KLOFE) & Institute of Advanced Materials (IAM), Jiangsu National Synergetic Innovation Center for Advanced Materials (SICAM), Nanjing Tech University (Nanjing Tech), Nanjing 211816, P. R. China
[d] Department of Materials and Environmental Chemistry (MMK), Stockholm University, Svante Arrheniusväg 16 C, 10691 Stockholm, Sweden
[e] Department of Chemistry and Biomolecular Science, Clarkson University, Potsdam NY 13699, USA

*Corresponding author. E-mail address: jiayin.yuan@mmk.su.se

**Abstract**: Porous heteroatom-doped carbons are desirable for catalytic reactions due to their tunable physicochemical properties, low cost and metal-free nature. Herein, we introduce a facile, general bottom-up strategy, so-called "cooking carbon in a solid salt", to prepare hierarchically porous heteroatom-doped carbon foams by using poly(ionic liquid) as precursor and a common inorganic salt NaCl as structural template. The obtained carbon foams bear micro-/meso-/macropores, large specific surface area and rich nitrogen dopant. The combination of these favorable features facilitates the catalytic degradation of aqueous organic pollutants by persulfate under visible light irradiation, in which they prevail over the state-of-the-art metal-/carbon-based catalysts.

## 1. Introduction

Emerging as an important class of multifunctional materials, porous heteroatom-doped carbons have captured ever-increasing research interest across the fields of biology, materials science, chemistry, physics and energy [1-6]. The incorporation of heteroatom functionalities into a carbon framework is expected to modify its electronic structure, Fermi level and surface basic/coordinating properties [7-11]. As such, it may amplify materials performance in a broad spectrum of applications [12-18], particularly in catalytic degradation of organic pollutants [19]. For a long time, organic pollutants in wastewater have been one of the major concerns in our society on the path to urbanization and industrialization [20-23]. Advanced

Oxidation Process (AOP) is the method of choice in terms of full decomposition of target organic pollutants. In this regard, sulfate radicals ($SO_4^{\bullet-}$) are more selective than hydroxyl radicals ($^{\bullet}OH$) and achieve better mineralization [24]. They can be generated through activation of persulfate or peroxymonosulfate by many methods such as using UV, transition metal and metal oxides [25]. Heteroatom-doped carbon in comparison to transition metal catalysts is of low-cost and free from toxic metal leaching problem, and has been reported for the persulfate activation [26, 27]. Nevertheless, fabricating new highly efficient and cost-effective metal-free carbocatalysts in replacement of precious metal-based catalysts is actively pursued for green chemical processes and practical usages but remains a grand challenge.

Generally, the catalytic behavior of carbocatalyst is closely related to hybridization of nanocarbon of distinct electronic structure, tunable electron feature by elemental doping and pore architecture if it is porous. The coexistence of micro-/meso-/macroporous structures advantageously in combination with a large specific surface area will improve the catalytic performance [28, 29], since macropores/large mesopores provide the transport highways for reactants and micropores/small mesopores guarantee the abundance of heteroatom-doped active sites on the surface [30-33]. The routine synthesis of such porous carbon popularly involves physical and/or chemical activations of existent heteroatom-doped carbon [34, 35], carbonization of heteroatom-rich precursor with template such as silica [36, 37] and others [38], or thermal doping of pre-formed porous carbon by reactive gas [1]. These unremitting efforts however suffer from the usage of harsh acids and/or alkalis that raises concerns on equipment corrosion, low production yield, and/or tedious and energy-intensive procedure. As a sustainable approach, mixtures of inorganic salts in molten state have recently been utilized as high-temperature reaction media and porogens to produce porous carbons [39, 40], in which micropores and small mesopores are preferably formed in correspondence to the sizes of salt clusters and their percolation structures. However, macropores and large mesopores are often poorly developed in the obtained carbon products.

In this contribution, for the first time we applied a chemically inert inorganic salt to synthesize heteroatom-doped carbon foam with combined micro-/meso-/macropores. Although by virtue of crystal surface, salts in a solid form have been reported in the fabrication of sheet-like metal oxide hybrids [41-44] such as $MoO_3$ and MoN nanosheets, they have not yet been exploited for porous carbon foam production. The high melting point of certain inorganic salts and their easy removal by post-carbonization aqueous washing enable to readily build up porous carbons bearing a myriad of macropores and large mesopores, which may combine together with the intrinsic micropores and small mesopores generated by some polymeric precursors. One of such precursors can be some poly(ionic liquid)s (PILs) [45, 46], which *via* a self-templating mechanism may produce abundant micro-/mesopores with rich heteroatom dopants (such as nitrogen) in the final carbon product without activation processes or externally added porogens, thus superior to traditional polymeric carbon precursors such as polypyrrole [34]. Combining the aforementioned aspects, functional porous carbon foams with hierarchical micro-, meso- and macropores and rich heteroatom dopants are readily produced on a large scale and work as metal-free carbocatalysts, as exemplified here by catalytic degradation of organic pollutants in the presence of persulfate under visible light irradiation.

## 2. Materials and methods
## 2.1. Materials

1-Vinylimidazole, bromoacetonitrile, bis(trifluoromethane sulfonyl)imide lithium salt (LiTf$_2$N), 2,2'-azobis(2-methylpropionitrile) (AIBN), NaCl, KCl, Na$_2$SO$_4$, K$_2$S$_2$O$_8$, Rhodamine B (RB), K$_2$S$_2$O$_3$, ethylenediaminetetraacetic acid disodium salt (EDTA), *tert*-butanol and *p*-benzoquinone were bought from commercial chemical companies such as Alfa and used without further purifications. Fullerene (C60), graphene (G) and graphene oxide (GO) were kindly provided by Nanjing XFNANO Materials Tech Co., Ltd (China). Multi-walled carbon nanotube (CNT) with a diameter of 35–60 nm and a length of ca. 30 μm was purchased from Chengdu Institute of Organic Chemistry (China). CNT was ball-milled at 400 rpm for 6 h before being used, and the length was reduced to 0.3–1.0 μm. Carbon nitride (*g*-C$_3$N$_4$, simplified as C$_3$N$_4$) was prepared by applying urea as precursor at 550 °C in air for 4 h at the heating rate of 4 °C min$^{-1}$. Porous carbon fiber (PCF) with the average diameter of ca. 25.4 μm and the length of 4 to 6 cm was kindly provided by Anshan Sinocarb Carbon Fibers Co., Ltd (China). Acetylene carbon (AC) was purchased from Linzi Qishun Chemical Co. (China). Nitrogen-doped carbon foam (NCF) was prepared by the carbonization of commercial melamine foam (trademark Basotect® W, BASF, Germany) at 750 °C for 1 h at the heating rate of 5 °C min$^{-1}$ under nitrogen atmosphere. All other chemicals were of analytical grade.

## 2.2. Synthesis of PILMI as the N-rich precursor

As depicted in Fig. S1, poly[3-cyanomethyl-1-vinylimidazolium bis(trifluoromethane sulfonyl)imide] (termed "PCMVImTf$_2$N", simplified as "PILMI") was synthesized *via* anion exchange with poly(3-cyanomethyl-1-vinylimidazolium bromide) (termed "PCMVImBr") using LiTf$_2$N salt in aqueous solution according to our previous method [47, 48]. The gel permeation chromatography (GPC) trace of PCMVImBr is shown in Fig. S2. The proton nuclear magnetic resonance ($^1$H NMR) spectrum of PILMI is displayed in Fig. S3.

## 2.3. Preparation of porous nitrogen-doped carbon foam (PNCF) using PILMI as precursor

Firstly, 10.0 g of NaCl was dissolved in 35 mL of water. Meanwhile, a defined amount of PILMI was dissolved in DMF. The NaCl aqueous solution and the PILMI solution in DMF were mixed together at a NaCl/PILMI mass ratio of 10 under vigorous agitation for 30 min, followed by solvent evaporation at 100 °C to obtain light white NaCl/PILMI composite. Afterwards, the composite was heated at 550 °C for 1 h under nitrogen atmosphere, and subsequently calcined at 750 °C for 1 h to yield NaCl/N-C composite. The heating rate was kept as 10 °C min$^{-1}$. After cooling down to room temperature, the hierarchically porous nitrogen-doped carbon foam (PNCF) was washed by water and dried. The carbon product prepared from the carbonization of PILMI in the absence of NaCl template under the similar fabrication process was denoted as "C-PILMI". Besides, irregular carbon with few macroscopic pores and randomly distributed carbon nanosheets were prepared by the similar way at a NaCl/PILMI mass ratio of 2 and 50, respectively.

### 2.4. Preparation of PNCF using KCl or Na$_2$SO$_4$ as template

Similar to PNCF at Section 2.3, different PNCFs were prepared by using PILMI as the precusor and KCl or Na$_2$SO$_4$ as solid salt template at 725 or 750 °C.

### 2.5. Characterization

GPC measurement was conducted at 25 °C on a NOVEMA-column with a mixture solution of 80% acetate buffer and 20% methanol as eluent (flow rate = 1.0 mL min$^{-1}$, PEO standards using RI detector-Optilab-DSP-Interferometric Refractometer). $^1$H NMR measurement using DMSO-$d_6$ as the solvent was carried out at room temperature using a Bruker DPX-400 spectrometer operating at 400 MHz. Scanning electron microscopy (SEM) measurements were carried out on a LEO 1550-Gemini electron microscope (acceleration voltage = 3 kV). Transmission electron microcopy (TEM) measurements were performed using a Zeiss EM 912 (acceleration voltage = 120 kV). High-resolution TEM (HRTEM) measurement was carried out on a FEI Tecnai G2 S-Twin transmission electron microscope operating at 200 kV. Combustion elemental analyses were done with a varioMicro elemental analysis instrument from Elementar Analysensysteme. Energy dispersive X-ray (EDX) maps were taken on SEM carrying an EDX spectrometer. The surface element composition was characterized by means of X-ray photoelectron spectroscopy (XPS) carried out on a VG ESCALAB MK II spectrometer using an Al K$\alpha$ exciting radiation from an X-ray source operated at 10.0 kV and 10 mA. N$_2$ adsorption/desorption experiments were performed with a Quantachrome Autosorb and Quadrasorb at 77 K, and the data were analyzed by using Quantachrome software. The specific surface area was calculated by using Brunauer-Emmett-Teller (BET) equation. The pore volume was calculated by using $t$-method. The pore size distribution was obtained by applying the quench solid density functional theory (QSDFT) on the adsorption branch and assuming slit-like geometry on carbon material kernel. The samples were degassed at 150 °C for 20 h before measurements. The mercury porosimetry experiment was conducted to measure the pore size distribution of PNCF with an Autopore III device (Micromeritics, USA) according to DIN 66133. X-ray diffraction (XRD) pattern was recorded on a Bruker D8 diffractometer using Cu K$\alpha$ radiation ($\lambda$ = 0.154 nm) and a scintillation counter. Raman spectrum was collected by using a confocal Raman microscope ($\alpha$300; WITec, Ulm, Germany) equipped with a 532 nm laser.

### 2.6. Catalytic degradation of RB under visible light irradiation

The degradation experiments of RB were carried out at 25 °C. In a typical run, 3.0 mg of catalyst such as PNCF was dispersed in 50 mL of RB solution (20 mg L$^{-1}$). The suspension was stirred in dark for 15 min to achieve the adsorption/desorption equilibrium between the solution and catalyst. Then, a specified amount of persulfate (12.5 mg) was added into the solution to initiate the degradation reaction under visible light irradiation using a 50 W white LED array (Bridgelux BXRA-50C5300; $\lambda$ > 410 nm). Thus the initial concentrations of the catalyst and persulfate were 60 and 250 mg L$^{-1}$, respectively. After the reaction started, a drop of solution sample (0.2 mL) was taken at given time intervals, quenched immediately by adding K$_2$S$_2$O$_3$ solution, diluted, and centrifuged at 9000 rpm for 2 min to remove catalyst for the later analysis of RB concentration. The absorbance of RB solution was measured by UV/Vis/NIR spectrophotometer (Lambda 900) to monitor the absorbance at $\lambda_{max}$ = 554 ± 1 nm, corresponding to the maximum absorbance. The concentration of RB solution was determined by linear regression equation obtained by plotting the calibration curve for RB over a range of concentrations.

For all kinetic experiments tested under different conditions, the degradation efficiency of RB was fitted with the pseudo first-order kinetic equation as follows:

$$\ln\frac{C_t}{C_0} = -k*t \tag{1}$$

where $C_0$ and $C_t$ are the concentration of RB at the time 0 and reaction time $t$, respectively; $k$ is the apparent rate constant (min$^{-1}$).

Four sets of quenching experiments by using ethanol, *tert*-butanol, EDTA and *p*-benzoquinone as quenching agents were performed to determine the radical species formed in the PNCF + persulfate system during the catalytic degradation of RB under visible light irradiation. Prior to the persulfate addition, the quenching agent was added into the reaction solution, and the final concentration of quenching agent was fixed as 2.0 mmol L$^{-1}$.

## 3. Results and discussion
### 3.1. Preparation and characterization

Fig. 1 illustrates the fabrication routine towards the targeted carbon foams *via* "cooking carbon in a solid salt" approach. A salt/PIL physical mixture was firstly prepared by dropwise addition of an aqueous salt solution into PIL solution in DMF under vigorous agitation, followed by thermal evaporation of solvents. The mixture was calcined below the melting point of the salt to convert PIL into carbon in the solid salt matrix. Upon salt removal by aqueous treatment, carbon foams bearing rich heteroatoms and pores of various size scales were obtained. Initially, a cationic PIL, namely poly[3-cyanomethyl-1-vinylimidazolium bis(trifluoromethane sulfonyl)imide] (termed "PILMI", chemical structure and characterization shown in Figs. S1−S3), and commercial NaCl (melting point: 801 °C) were selected as a model system to prepare hierarchically porous nitrogen-doped carbon foam (PNCF) at 750 °C.

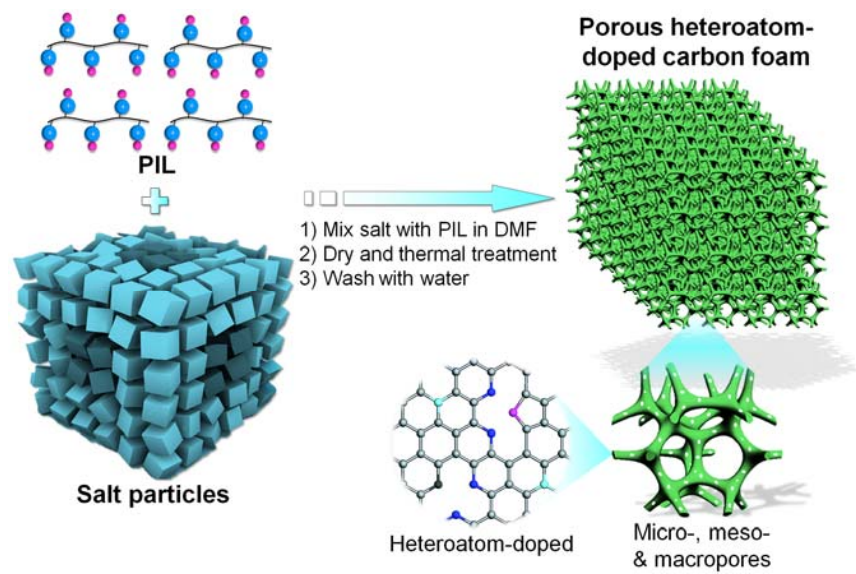

**Fig. 1.** General fabrication procedure of hierarchically porous heteroatom-doped carbon foams using PIL as precursor and chemically inert solid salt as template.

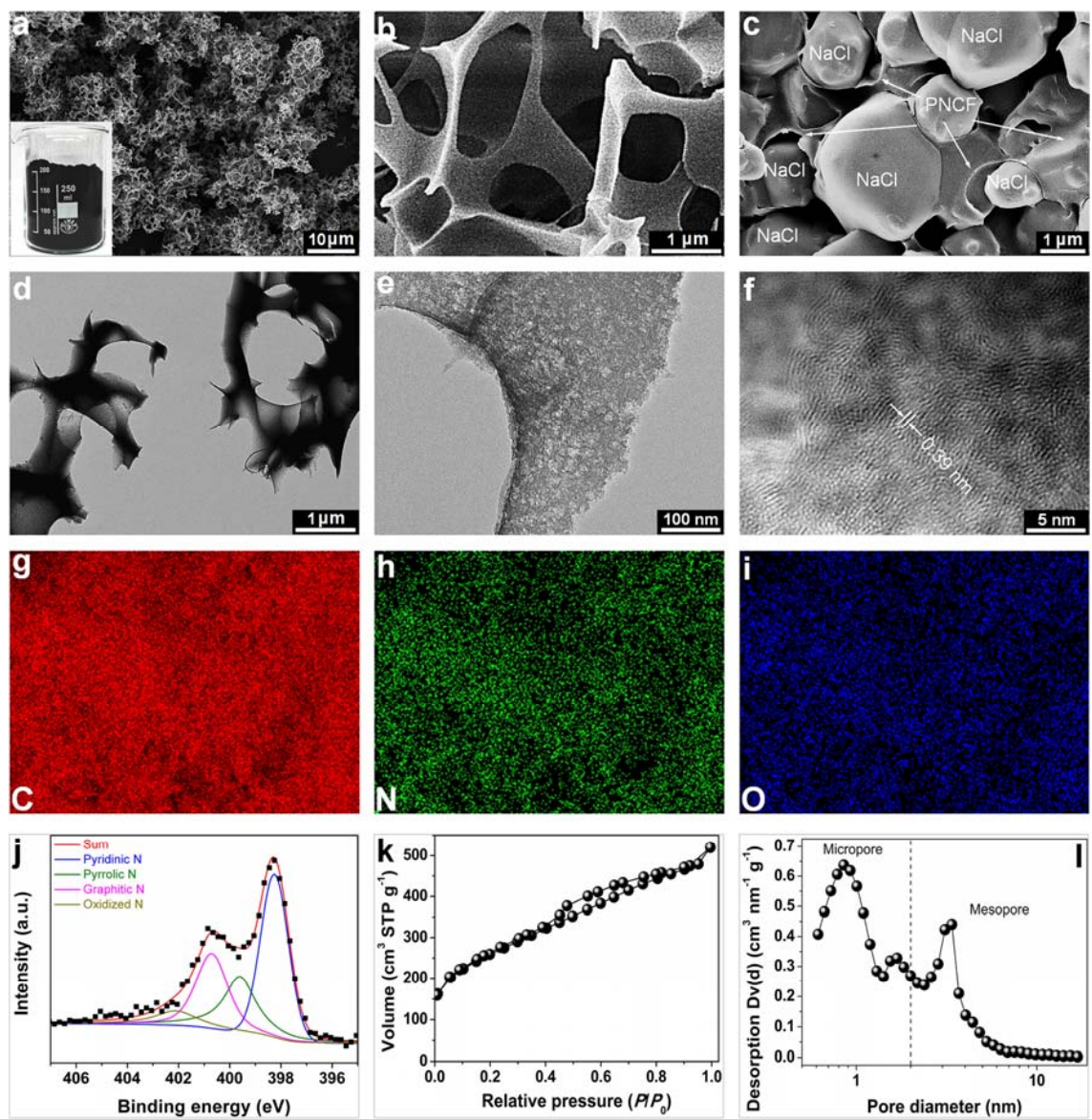

**Fig. 2.** (a and b) SEM imagesof PNCF, (c) SEM image of PNCF before NaCl removal, and (d and e) TEM images, (f) HRTEM image, (g–i) EDX maps according to (b), (j) high-resolution N 1s XPS curve, (k) N$_2$ adsorption/desorption isotherms at 77 K, and (l) the pore size distribution plot of PNCF. The inset in (a) shows the photograph of 12 g of PNCF.

SEM images of the as-synthesized PNCF at a NaCl/PILMI mass ratio of 10 (Figs. 2a, 2b and S4) reveal an open-cell, interconnected cellular network. Pores in a wide size range from ca. 0.5 µm to several microns are immediately identified. The pore walls are overall below 0.3 µm in thickness. In comparison with that of the non-washed sample (Figs. 2c and S5), it is easy to conclude that these micron-scale channels are produced from the elimination of the NaCl template. As a supportive proof, such micron-sized voids are absent in the carbon product prepared by carbonization of PILMI in the absence of NaCl (termed "C-PILMI", Fig. S6). To be mentioned is that the mass ratio of NaCl to PILMI at 10 is crucial in the formation of PNCF with ideal porous morphology. For example, at a low NaCl/PILMI mass ratio of 2, the resultant irregular carbon exhibits dramatically less micron-scale pores (Fig. S7), while at a high NaCl/PILMI mass ratio of 50, randomly distributed carbon nanosheets are received. Unlike carbon foams produced previously from $SiO_2$ [49], $CaCO_3$ [50], and nickel foam [51] templates, which involved aggressive chemicals to remove templates that are hardly reused, NaCl is nontoxic and readily dissolved in water, and ca. 95% of the used NaCl can be recycled for the next carbon foam production in this work (Fig. S8).

The simple implementation of carbonization enables large-scale synthesis of PNCF, *e.g.*, 12 g in lab (Fig. 1a inset). TEM images of PNCF indicate the presence of abundant large mesopores inside the pore wall (Figs. 2d and 2e). HRTEM image (Fig. 2f) clearly shows the bent graphitic sheets, typical for nitrogen-doped carbons. The interplanar spacing is calculated by Fast Fourier transform to be 0.39 nm, 15% larger than that of native graphite (0.34 nm), which may result from the accommodation of the bulky lone electron pair of nitrogen atom in the 2D carbon framework. Combustion elemental analysis of PNCF reveals a high nitrogen content of 17.1 wt % (Table S1), and the elemental mapping based on EDX spectroscopy (Figs. 2g−2i) portrays the homogeneous distribution of nitrogen as well as carbon and oxygen. To elucidate the chemical status of nitrogen atoms in PNCF, high-resolution N 1s spectrum is illuminated by XPS and deconvoluted into four individual peaks of pyridinic, pyrrolic, graphitic and oxidized nitrogen [52] (Figs. 2j and S9). Pyridinic nitrogen (43.9%) and graphitic nitrogen (26.4%) are the two dominant types.

The textural property of PNCF was investigated by $N_2$ physisorption isotherm (Figs. 2k and 2l). The significant $N_2$ uptake takes place in the low-pressure range ($P/P_0 < 0.1$), testifying the presence of a large number of micropores. The detectable type-H4 hysteresis loop at a relative pressure between 0.4 and 0.8 corresponds to the filling and emptying of mesopores by capillary condensation. The Brunauer-Emmett-Teller specific surface area ($S_{BET}$) and pore volume are determined to be 931 $m^2$ $g^{−1}$ and 0.68 $cm^3$ $g^{−1}$, respectively. By contrast, the $S_{BET}$ and pore volume of C-PILMI are merely 708 $m^2$ $g^{−1}$ and 0.38 $cm^3$ $g^{−1}$, respectively (Fig. S10, Table S1). Apparently, PNCF possesses 32% more $S_{BET}$ than C-PILMI and significantly 80% more pore volume in the micro-/mesopore size range, not to mention the extra micron-sized channels observed in SEM characterization (Figs. 2a and 2b) and mercury intrusion porosimetry (Fig. S11), which is a proof of advantage to employ the concept of solid salt template.

XRD and Raman measurements were conducted to figure out the phase composition of PNCF. Two weak diffraction peaks centered on $2\theta = 25°$ (002) and 45° (100) are observed, suggesting the turbostratic character of PNCF (Fig. 3a). The absence of other peaks indicates the complete removal of NaCl. Raman spectrum shows sharp *D* and *G* peaks at 1340 and 1570 $cm^{−1}$, respectively (Fig. 3b). The *G* peak arises from in-plane vibration of $sp^2$ C–C network and is a measure of structural order in carbon [53]. Disorder as well as bonding of heteroatoms to carbon distinctly contributes to the intensity of *D* peak. The small value of $I_D/I_G \sim 0.5$ reflects the presence of plenty of disorders, edges and boundaries of graphitic domains, typical for highly porous carbons. It is worth noting that the solid salt approach can employ a wide range of commercially available inorganic salts bearing high melting points such as potassium chloride and sodium sulfate. Both salts have been successfully exploited to template the growth of carbon foams (Figs. S12 and S13).

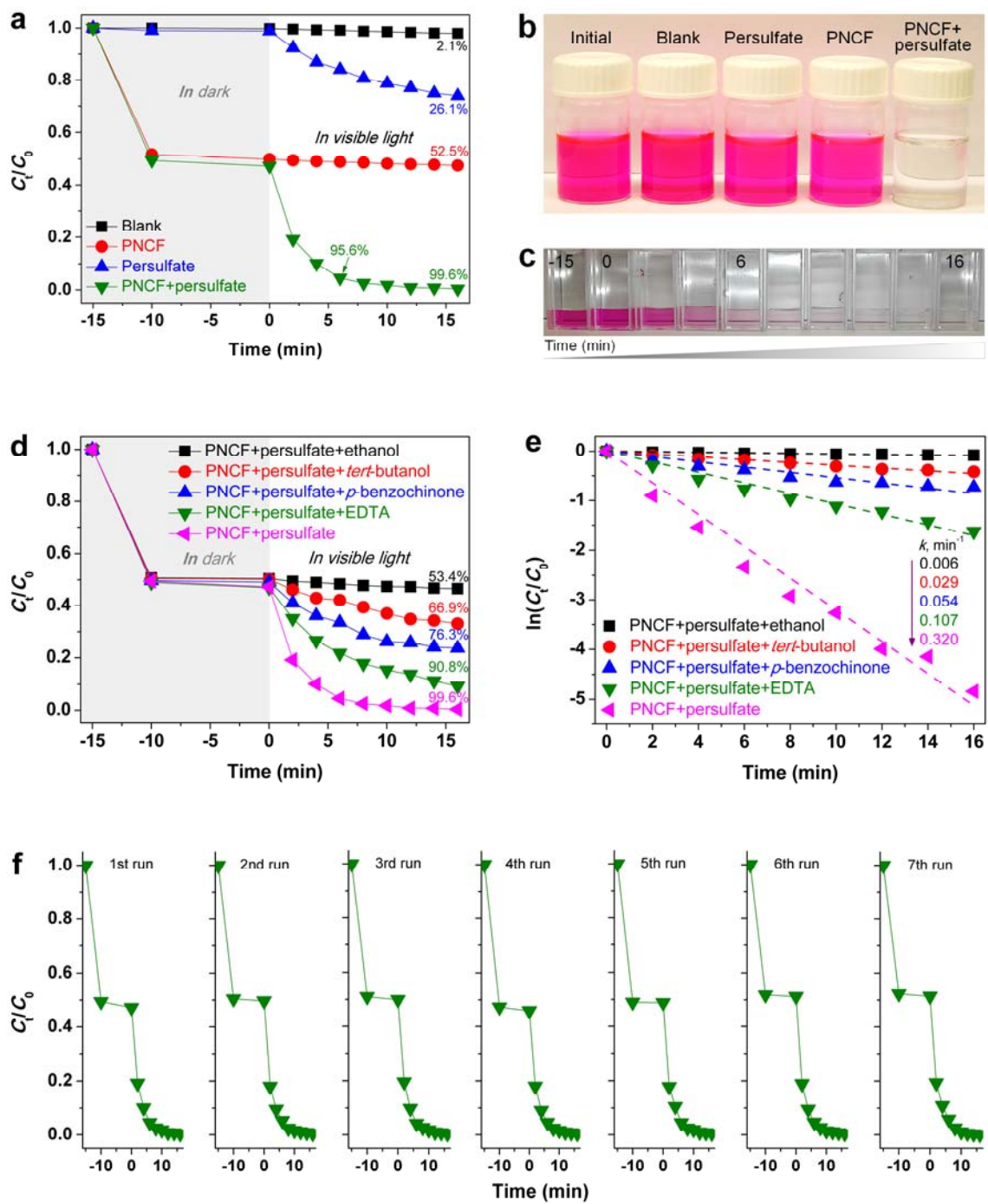

**Fig. 3.** (a) Time-dependent catalytic degradation of RB (20 mg L$^{-1}$) in aqueous solution containing no catalysts (black), PNCF (60 mg L$^{-1}$, red), persulfate (250 mg L$^{-1}$, blue), and PNCF + persulfate (60 and 250 mg L$^{-1}$, respectively, green). (b) Photographs of RB solution in different systems under visible light after 16 min. (c) Photographs of RB solution in the PNCF + persulfate system at different irradiation time. (d and e) Effect of radical scavenger on the removal efficiency of RB. (f) Multiple cycles of catalytic degradation profiles of RB in the PNCF + persulfate system.

## 3.2. Photocatalytic degradation of organic pollutant

The activation of persulfate (or peroxymonosulfate) to generate $SO_4^{•-}$ radicals is an important industrial reaction to degrade organic pollutants in wastewater. In spite of the development of a number of homogeneous and heterogeneous catalysts that promote this activation [54], an ideal catalyst, which is of low-cost, metal-free, environment-benign, highly active under mild conditions and recyclable, is still eagerly pursued. In this study, we explored the catalytic degradation of organic pollutants, here Rhodamine B (RB) as a model organic dye, in the presence of PNCF and persulfate as a co-catalyst system under visible light irradiation.

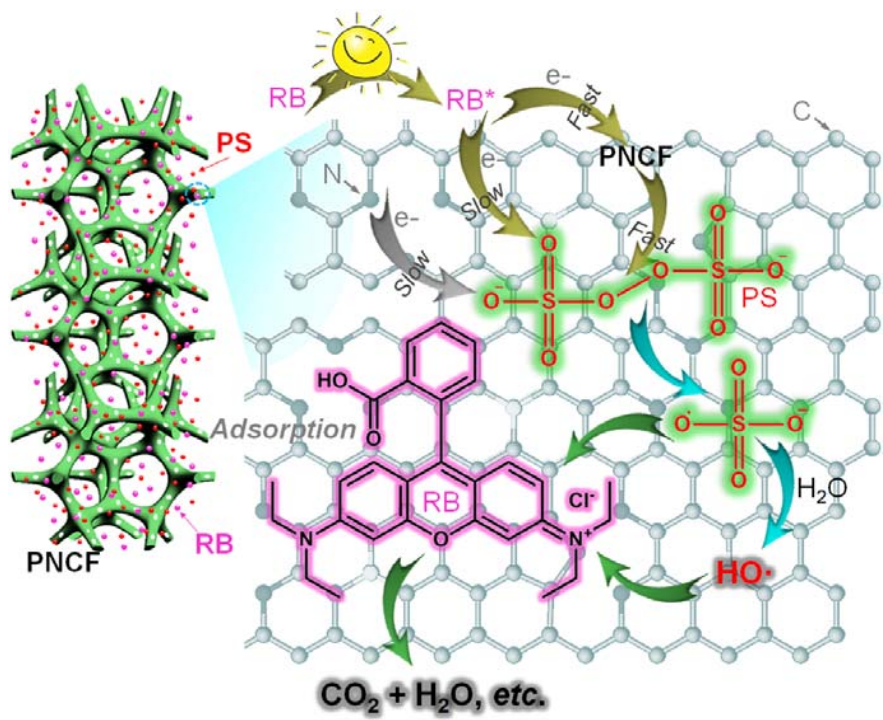

**Fig. 4.** Schematic catalytic degradation mechanism of RB in the PNCF + persulfate system.

Fig. 4a presents the removal efficiency of RB in different environments under visible light irradiation. In this figure (and other similar figures beneath), the RB degradation starts at a time point defined as "0" right after persulfate addition. Before this time point "0", the removal of RB from the solution phase is purely adsorption-driven by PNCF. The total removal of RB thus consists of the contributions from adsorption by PNCF and oxidative degradation by persulfate. For the sake of clarity, firstly a series of control experiments were conducted. Under visible light irradiation for 16 min, the degradation efficiency of RB (merely 2.1%) is negligible , which increases to 26.1% after persulfate addition. PNCF itself can store 52.5% of RB in its micro-/mesopores by adsorption mechanism (Fig. 4b). By contrast, the combination of persulfate and PNCF under visible light irradiation readily removes 95.6% of RB within 6 min, which reaches 99.6% in the next 10 min (Fig. 4c).

The mathematic data fitting in the oxidation stage (Fig. S14) gives an apparent rate constant ($k$) of 0.320 min$^{-1}$ for the catalytic degradation of RB in the PNCF + persulfate system, which is 14 and 68 times higher than that of the persulfate system (0.021 min$^{-1}$) and the PNCF system (0.003 min$^{-1}$), respectively. Thereupon the synergy between PNCF and persulfate does exist in the catalytic degradation of RB. Furthermore, to assess the reusability, PNCF was separated by centrifuge, washed and dried for its next use. The RB degradation rate of the reclaimed PNCF is similar to that of the fresh one (Fig. 4f), and its degradation efficiency reaches 99.5% after 7 cycles, proving that PNCF is stable for repeated uses.

To clarify the origin of the high activity of the PNCF + persulfate system, we performed radical trapping experiments by using different radical scavengers (Figs. 4d and 4e). When ethanol (the scavengers of both SO$_4$·$^{-}$ and ·OH radicals) or *tert*-butanol (the scavenger of ·OH radical) is added, the catalytic activity is inhibited significantly. In contrast, when *p*-benzoquinone (the scavenger of O$_2$·$^{-}$ radical) or disodium ethylenediaminetetraacetate (EDTA, the scavenger of h$^+$) is introduced, the process is less affected. Thereby, SO$_4$·$^{-}$ and ·OH radicals are the main reactive species.

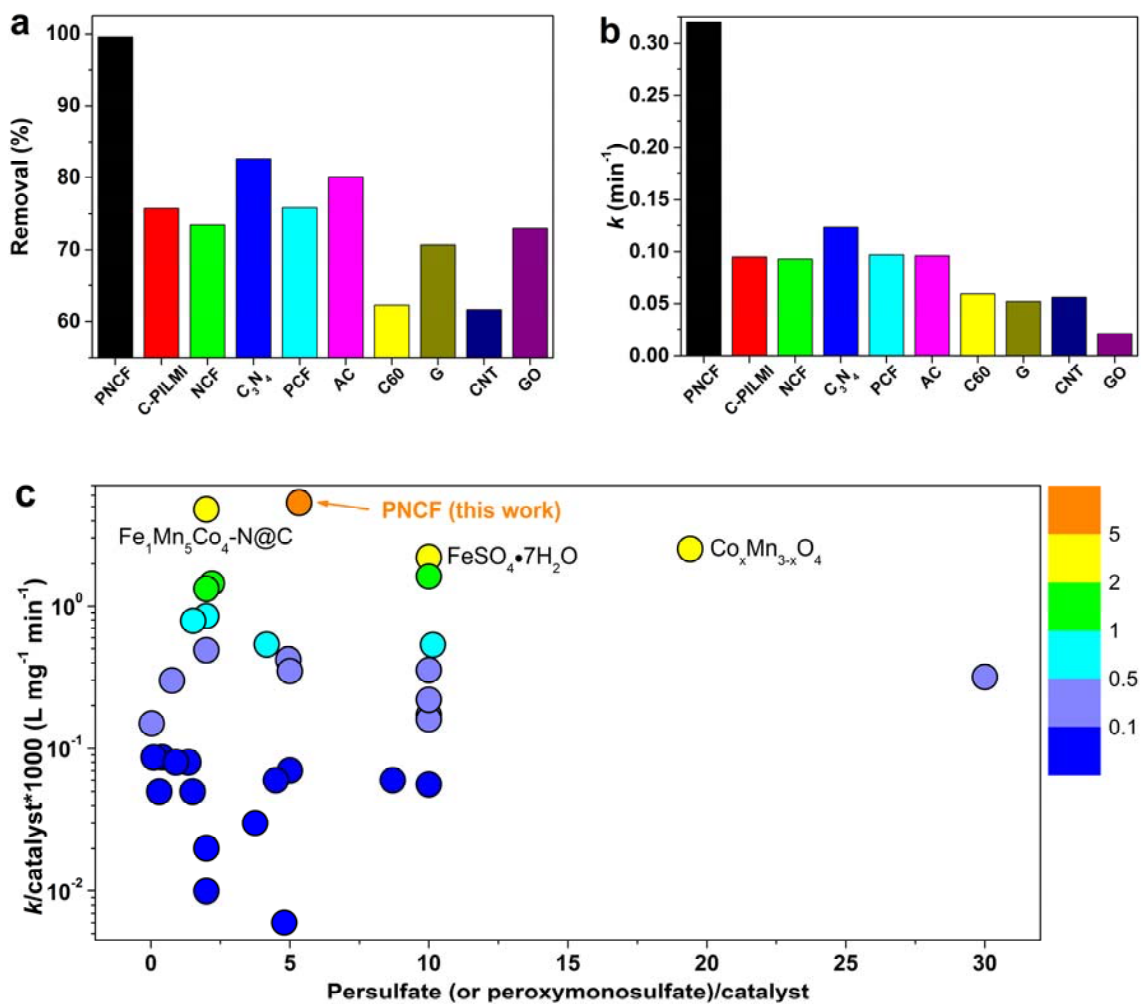

**Fig. 5.** Comparisons of (a) RB removal efficiency and (b) apparent rate constant (*k*) of MB for PNCF with traditional carbonaceous materials. (c) Comparisons of *k*/catalyst*1000 and persulfate (or peroxymonosulfate)/catalyst for PNCF with previously reported carbon-/metal-based catalysts (Table S4).

Persulfate activation relies on electron transfer from catalyst to break up the O–O bond of persulfate, thus producing $SO_4^{\cdot-}$ radical [55]. Since RB is the visible light absorbing species, it is expected that the excited state of RB (RB*) will inject an electron into the O–O bond of persulfate where it is captured to produce $SO_4^{\cdot-}$ radical (Fig. 5). A portion of formed $SO_4^{\cdot-}$ radical catalyzes the decomposition of $H_2O$ to form $\cdot OH$ radical [25]. These as-generated radicals take part in the degradation of RB into $CO_2$, $H_2O$, *etc*. It explains the decomposition of RB in the presence of persulfate alone (26.1% for 16 min). In the presence of PNCF, this process runs much more efficiently, as observed. On one hand, the molecular sizes of RB and persulfate are ca. 0.6 and 1.6 nm, respectively, matching well with the micropores and small mesopores in PNCF, and thereby are significantly enriched in these pores. Meanwhile hydrogen bonding interaction (*e.g.*, the nitrogen atom of pyridinic C–N group can serve as a hydrogen-bonding acceptor), electrostatic interaction (Fig. S15) and $\pi$-$\pi$ interaction (since RB is a planar molecule) between RB/persulfate and PNCF facilitate their adsorptions to PNCF. This enrichment process is prerequisite for promoting the electron transfer from RB* to persulfate. On the other hand, due to the $sp^2$ carbon in graphitic layers, the zigzag edges with unconfined $\pi$ electrons [56, 57], and the doping nitrogen atom that bears higher electronegativity (3.04 eV) than carbon (2.55 eV) as well as higher electron density, PNCF itself presents potential to transport the abundant free-flowing electrons to persulfate, as proved by the slow but continuous decomposition of RB in dark (Fig. S16). That is to say, PNCF not only transfers the excited electrons from RB* to persulfate, but also provides some actives for facilitating the activation of persulfate.

Additionally, we compared the catalytic performance of PNCF with other carbonaceous materials (Figs. S6 and S17, Table S3), including C-PILMI, nitrogen-doped carbon foam (NCF), carbon nitride ($C_3N_4$), porous carbon fiber (PCF), fullerene (C60), CNT and more. The results are summarized in Figs. 6a, 6b, S18 and S19. The RB degradation rate of PNCF is overall 3–10 times that of other carbocatalysts under the same experimental conditions. For instance, microporous C-PILMI in spite of high nitrogen content (14.5 wt %) and $S_{BET}$ (708 $m^2$ $g^{-1}$) shows a low pre-adsorption capacity of RB (5%) that lows down the overall degradation efficiency of RB. Similar phenomena are observed in the case of microporous PCF that exhibits $S_{BET}$ of 801 $m^2$ $g^{-1}$. Comparatively, GO containing rich functional groups displays a high pre-adsorption capacity of RB (63%), but its nitrogen-free nature leads to remarkably lower degradation rate of RB. In the cases of NCF and $C_3N_4$, both bear a high nitrogen dopant (26.1 and 60.6 wt %, respectively); unfortunately their poor nanoporous structure ($S_{BET}$ < 100 $m^2$ $g^{-1}$) leads to a very low pre-adsorption capacity of RB (< 2%) and thus inferior degradation efficiency. When acetylene carbon (AC), graphene (G), C60 and CNT are tested, the degradation rate of RB is generally lower, as a result of nitrogen-free character and/or scarcity of appropriate pores. It should be noted that although NCF bears a greatly lower $S_{BET}$ of 92 $m^2$ $g^{-1}$ than PCF, the apparent rate constant of RB for NCF is comparable to that for PCF, confirming the indispensable role of nitrogen dopant in the persulfate activation.

The results above demonstrate the synergistic effect of rich nitrogen dopant and hierarchical pore structure of the PNCF catalyst on the catalytic degradation of pollutant. More importantly, PNCF outperforms the state-of-the-art carbon-/metal-based catalysts (*e.g.*,

$Fe_1Mn_5Co_4$-N@C [58], Fig. 6c, Table S4) in terms of the degradation rate of pollutant (that is, $k$/catalyst*1000) and the activation efficiency of persulfate or peroxymonosulfate (that is, persulfate (or peroxymonosulfate)/catalyst).

## 4. Conclusions

In summary, we have reported a universal approach of conducting carbonization in a solid salt matrix that easily accesses hierarchically porous heteroatom-doped carbon foams and their hybrids. The commercial salts as scaffold simplify the synthetic procedure and are recyclable. Thanks to the abundant heteroatom dopants and a broad size range of pores, in the presence of persulfate they work excellently in the catalytic degradation of organic pollutants, exemplified here by Rhodamine B, under visible light irradiation, and outperform the state-of-the-state metal-based and carbonaceous catalysts. Considering the easy accessibility of various solid salts and rich choices of polymeric precursors or additives, this synthetic approach displays in our opinion unprecedented versatility and addresses a variety of functional porous carbons for applications in catalysis, environment remediation, energy production and storage, sensor and more. We expect that in the future commercial water-soluble, bio-based PILs can be produced and used to prepare such carbon foams in order to avoid the usage of organic solvents and make the fabrication process of PIL-derived carbon foams more cost-effective and benign.

## Acknowledgements
J.Y. thanks the Clarkson University for a start-up grant, the European Research Council Starting Grant (639720-NAPOLI), and the Strategic Support fund (project number 4601201) from Stockholm University.

*Supplementary Material*

**Table S1.** Yields and element compositions of C-PILMI and PNCF.

| Sample | Yield [a] (wt %) | C [b] (wt %) | N [b] (wt %) | H [b] (wt %) | S [b] (wt %) | O [c] (wt %) |
|---|---|---|---|---|---|---|
| C-PILMI | 21.7 | 64.3 | 14.5 | 2.2 | 0.02 | 19.0 |
| PNCF [d] | 30.3 | 64.2 | 17.1 | 2.0 | 0.02 | 16.7 |
| PNCF [e] | 31.2 | 64.4 | 16.9 | 1.9 | 0.02 | 16.8 |
| PNCF [f] | 30.5 | 64.8 | 17.2 | 2.0 | 0.03 | 16.0 |

[a] Calculated by the mass ratio of PILMI-derived carbon fraction to PILMI in the composite. [b] Measured by combustion element analyses. [c] Calculated by the difference. [d] Prepared by using NaCl (melting point = 801 °C) particle as the solid salt template at 750 °C. [e] Prepared by using KCl (melting point = 770 °C) particle as the solid salt template at 725 °C. [f] Prepared by using $Na_2SO_4$ (melting point = 884 °C) particle as the solid salt template at 750 °C.

**Table S2.** Textural parameters of C-PILMI and PNCF.

| Sample | $S_{BET}$ [a] ($m^2\ g^{-1}$) | $S_{micro}$ [b] ($m^2\ g^{-1}$) | $V_{total}$ [c] ($cm^3\ g^{-1}$) | $V_{micro}$ [d] ($cm^3\ g^{-1}$) | $D_A$ [e] (nm) |
|---|---|---|---|---|---|
| C-PILMI | 708 | 524 | 0.38 | 0.28 | 1.2 |
| PNCF [f] | 931 | 577 | 0.68 | 0.28 | 1.5 |
| PNCF [g] | 952 | 541 | 0.53 | 0.27 | 1.5 |
| PNCF [h] | 907 | 500 | 0.51 | 0.25 | 1.5 |

[a] BET specific surface area. [b] Specific surface area of micropore. [c] Pore volume. [d] Volume of micropore. [e] Average size of pore. [f] Prepared by using NaCl particle as the solid salt template. [g] Prepared by using KCl particle as the solid salt template. [h] Prepared by using $Na_2SO_4$ particle as the solid salt template.

**Table S3.** Element compositions and textural parameters of the reference carbon-based materials.

| Sample | C [a] (wt %) | N [a] (wt %) | H [a] (wt %) | O [b] (wt %) | $S_{BET}$ [c] ($m^2\ g^{-1}$) | $S_{micro}$ [d] ($m^2\ g^{-1}$) | $V_{total}$ [e] ($cm^3\ g^{-1}$) | $V_{micro}$ [f] ($cm^3\ g^{-1}$) |
|---|---|---|---|---|---|---|---|---|
| NCF | 55.2 | 26.1 | 1.5 | 17.2 | 92 | 71 | 0.02 | 0.02 |
| $C_3N_4$ | 35.1 | 60.6 | 2.0 | 2.3 | 97 | 30 | 0.10 | 0.02 |
| PCF | 84.3 | 0 | 1.4 | 14.3 | 801 | 750 | 0.40 | 0.38 |
| AC | 96.2 | 0 | 0.9 | 2.9 | 240 | 23 | 0.40 | 0.01 |
| C60 | 93.4 | 0 | 1.0 | 5.6 | 229 | 112 | 0.32 | 0.05 |
| G | 97.9 | 0 | 0.9 | 1.2 | 219 | - | 0.08 | - |
| CNT | 98.5 | 0 | 0.8 | 0.7 | 58 | - | 0.21 | - |
| GO | 45.9 | 0 | 2.8 | 51.3 | 85 | - | 0.01 | - |

[a] Measured by combustion element analyses. [b] Calculated by the difference. [c] BET specific surface area. [d] Specific surface area of micropore. [e] Pore volume. [f] Volume of micropore.

**Table S4.** Comparisons of the apparent rate constant ($k$), the degradation rate of pollutant, and the activation efficiency of persulfate by PNCF with those of other carbon- and metal-based catalysts from the previous work.

| Entry | Light, heat or other [a] | Persulfate or peroxymonosulfate [b] | Concentration [c] (mg L$^{-1}$) | Catalyst [d] | Concentration [e] (mg L$^{-1}$) | $k$ [f] (min$^{-1}$) | Persulfate (or peroxymonosulfate)/catalyst [g] | $k$/catalyst*1000 [h] (L mg$^{-1}$ min$^{-1}$) | N [i] (wt %) | $S_{BET}$ [g] (m$^2$ g$^{-1}$) | Ref. |
|---|---|---|---|---|---|---|---|---|---|---|---|
| 1 | Visible light | Persulfate | 250 | PNCF | 60 | 0.320 | 4.17 | **5.33** | 17.1 | 931 | This work |
| 2 | - [k] | Peroxymonosulfate | 200 | Fe$_1$Mn$_5$Co$_4$-N@C | 100 | 0.48 | 2 | **4.80** | - | 66 | [1] |
| 3 | - | Persulfate | 54 | FeSO$_4$·7H$_2$O | 2.8 | 0.007 | 19.4 | **2.52** | - | - | [2] |
| 4 | - | Peroxymonosulfate | 200 | Co$_x$Mn$_{3-x}$O$_4$ | 20 | 0.044 | 10 | **2.20** | - | - | [3] |
| 5 | - | Peroxymonosulfate | 1000 | N,S-codoped CNT-COOH | 100 | 0.162 | 10 | **1.62** | 3.7 | 192 | [4] |
| 6 | - | Peroxymonosulfate | 657.5 | Co-MOFs | 300 | 0.433 | 2.19 | **1.44** | 1.5 | 1 | [5] |
| 7 | - | Peroxymonosulfate | 200 | Fe$_3$Co$_7$@C-650 | 100 | 0.132 | 2 | **1.32** | - | 103 | [6] |
| 8 | - | Peroxymonosulfate | 100 | Fe@C-BN6700 | 20 | 0.02 | 5 | **1.00** | - | - | [7] |
| 9 | - | Peroxymonosulfate | 200 | Cobalt powder | 100 | 0.085 | 2 | **0.85** | - | - | [6] |
| 10 | - | Peroxymonosulfate | 250 | Porous reduced graphene oxide | 60 | 0.032 | 4.17 | **0.54** | 0 | 900 | [8] |
| 11 | - | Peroxymonosulfate | 3044 | SiO$_2$@Fe$_3$O$_4$/C | 300 | 0.16 | 10.15 | **0.53** | - | 248 | [9] |
| 12 | - | Peroxymonosulfate | 200 | Fe$_{0.8}$Co$_{2.2}$O$_4$ nanocage | 100 | 0.049 | 2 | **0.49** | - | - | [6] |
| 13 | - | Persulfate | 989.3 | Nanodiamond | 200 | 0.084 | 4.95 | **0.42** | 1.6 | 364 | [10] |
| 14 | - | Peroxymonosulfate | 2000 | N-doped graphene | 200 | 0.071 | 10 | **0.36** | 0.8 | 99 | [11] |
| 15 | - | Peroxymonosulfate | 500 | CuO-Co$_3$O$_4$@MnO$_2$ | 100 | 0.035 | 5 | **0.35** | - | 31.8 | [12] |
| 16 | - | Peroxymonosulfate | 1500 | CoFe/CoFe$_2$O$_4$ | 50 | 0.016 | 30 | **0.32** | - | - | [13] |
| 17 | - | Persulfate | 76.1 | Zero-valent iron | 100 | 0.03 | 0.76 | **0.30** | - | - | [14] |
| 18 | - | Peroxymonosulfate | 2000 | Fe/Fe$_3$C@NC | 200 | 0.043 | 10 | **0.22** | 1.8 | 160 | [15] |
| 19 | - | Peroxymonosulfate | 1000 | CuFe$_2$O$_4$ | 100 | 0.017 | 10 | **0.17** | - | - | [4] |
| 20 | - | Persulfate | 2000 | N-doped reduced GO | 200 | 0.032 | 10 | **0.16** | 7.4 | 96 | [16] |
| 21 | - | Peroxymonosulfate | 15.2 | CuFe$_2$O$_4$ | 500 | 0.075 | 0.03 | **0.15** | - | - | [17] |

| | | | | | | | | | | |
|---|---|---|---|---|---|---|---|---|---|---|
| 22 | - | Peroxymonosulfate | 200 | Graphene | 500 | 0.044 | 0.40 | **0.09** | 0 | - | [18] |
| 23 | - | Peroxymonosulfate | 152.2 | $Fe_2O_3$ | 1500 | 0.13 | 0.10 | **0.09** | - | 34 | [19] |
| 24 | - | Persulfate | 1351.6 | S/Fe composite | 1000 | 0.084 | 1.35 | **0.08** | - | - | [20] |
| 25 | - | Persulfate | 2500 | $Bi_2WO_6/Fe_3O_4$ | 500 | 0.035 | 5 | **0.07** | - | - | [21] |
| 26 | - | Peroxymonosulfate | 1000 | CNT | 100 | 0.006 | 10 | **0.06** | - | - | [4] |
| 27 | - | Persulfate | 8650 | Metal-organic frameworks MIL-53(Fe) | 1000 | 0.056 | 8.7 | **0.06** | - | 89.7 | [22] |
| 28 | - | Peroxymonosulfate | 152 | $NaHSO_3$ | 104 | 0.005 | 1.5 | **0.05** | - | - | [23] |
| 29 | - | Peroxymonosulfate | 609 | FeCit@ACFs | 2000 | 0.096 | 0.3 | **0.05** | - | - | [24] |
| 30 | - | Persulfate | 3000 | $\alpha$-FeOOH | 800 | 0.023 | 3.75 | **0.03** | - | - | [25] |
| 31 | - | Peroxymonosulfate | 200 | $Co_3O_4$ powder | 100 | 0.002 | 2 | **0.02** | - | - | [6] |
| 32 | - | Peroxymonosulfate | 200 | Fe powder | 100 | 0.001 | 2 | **0.01** | - | - | [6] |
| 33 | - | Persulfate | 9654 | Nanoscale iron | 2000 | 0.012 | 4.8 | **0.006** | - | - | [26] |
| 34 | UV | Persulfate | 1217.6 | - | - | 0.311 | - | - | - | - | [27] |
| 35 | UV | Persulfate | 7298.7 | - | - | 0.154 | - | - | - | - | [28] |
| 36 | UV | Persulfate | 510.9 | - | - | 0.033 | - | - | - | - | [29] |
| 37 | UV | Persulfate | 6758 | - | - | 0.022 | - | - | - | - | [30] |
| 38 | UV | Persulfate | 913.2 | - | - | 0.020 | - | - | - | - | [31] |
| 39 | $T = 80$ °C | Persulfate | 41.9 | - | - | 0.025 | - | - | - | - | [32] |
| 40 | $T = 70$ °C | Persulfate | 135.2 | - | - | 0.004 | - | - | - | - | [33] |
| 41 | $\gamma$-ray | Peroxymonosulfate | 21.3 | - | - | 0.005 | - | - | - | - | [34] |

[a] Under UV, visible light or other irradiations, or by heating. [b] In addition to persulfate, peroxymonosulfate is also widely used to produce $SO_4^{•-}$ in the Advanced Oxidation Processes (AOPs). [c] The concentration of persulfate or peroxymonosulfate. [d] Various catalysts are used in the activation of persulfate or peroxymonosulfate to produce $SO_4^{•-}$. [e] The concentration of catalyst. [f] The apparent rate constant ($k$) of pollutant. [g] The persulfate or peroxymonosulfate activation efficiency, that is to say, the concentration ratio of the persulfate or peroxymonosulfate to the catalyst; the low value of persulfate (or peroxymonosulfate)/catalyst means the need of a relatively low amount of persulfate or peroxymonosulfate when the content of catalyst keeps the same. [h] The degradation rate of pollutants catalyzed by catalyst, that is to say, the 1000 times of the ratio of the apparent rate constant to the catalyst concentration; the high degradation rate of pollutants suggests the higher

apparent rate constant of pollutants when the same amount of catalyst is used. $^i$ The nitrogen content of catalyst. $^j$ BET specific surface area of catalyst. $^k$ Not given.

It should be pointed out that the degradation rate of the pollutant catalyzed by PNCF (*i.e.*, $k$/catalyst*1000) is higher than that of other carbon- or metal-based catalysts from the previous work (entries 2–33). Besides, the PNCF + persulfate system under visible light irradiation in our work outperforms the UV-, heat-, or $\gamma$-ray-based systems (entries 34–41) in terms of higher apparent rate constant ($k$) of pollutant.

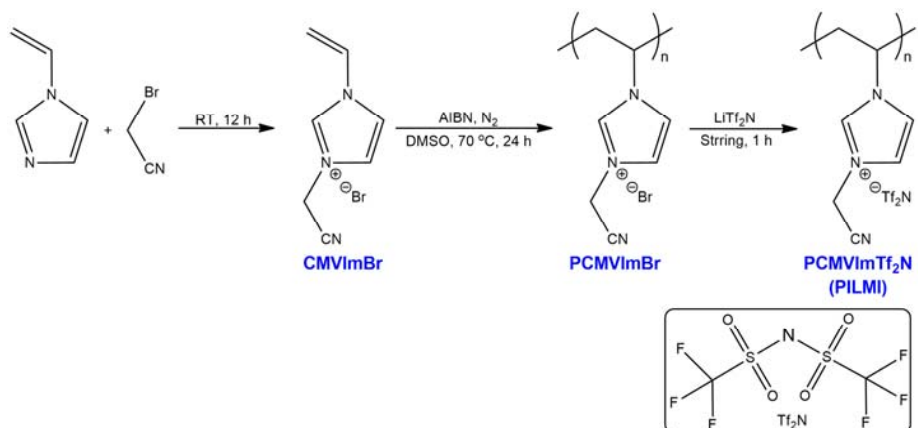

**Fig. S1.** Synthetic routes towards PILMI applied as precursor to prepare the hierarchically porous nitrogen-doped carbon foam in this work.

Poly(3-cyanomethyl-1-vinylimidazolium bromide) (termed "PCMVImBr") was synthesized according to our previous work [35, 36]. Briefly, 10.0 g of the monomer 3-cyanomethyl-1-vinyl imidazolium bromide (termed "CMVImBr", prepared from the reaction of 1-vinylimidazole and bromoacetonitrile at room temperature for 12 h), 200 mg of AIBN and 100 mL of DMSO were loaded into a 250 mL of Schlenk flask. The mixture was deoxygenated several times by a freeze-pump-thaw procedure. The reactor was then refilled with nitrogen and placed in an oil bath at 70 °C for 24 h. The mixture was exhaustively dialyzed against water for one week (the molecular weight cut-off of the dialysis bag is 8 kDa) and freeze-dried from water.

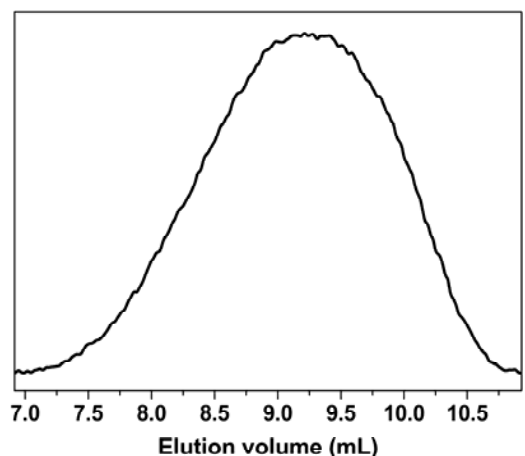

**Fig. S2.** GPC trace of PCMVImBr.

The apparent number-average molecular weight and polydispersity index value of PCMVImBr are measured to be $1.97 \times 10^5$ g mol$^{-1}$ and 2.67, respectively. PILMI is synthesized by anion exchange of PCMVImBr with LiTf$_2$N salt in aqueous solution. Therefore, the apparent number-average molecular weight of PILMI is $3.67 \times 10^5$ g mol$^{-1}$.

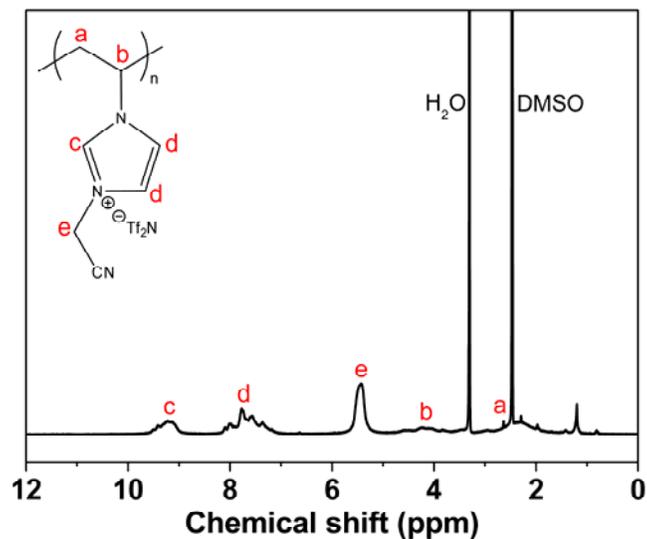

**Fig. S3.** $^1$H NMR spectra of PILMI using DMSO-$d_6$ as the solvent.

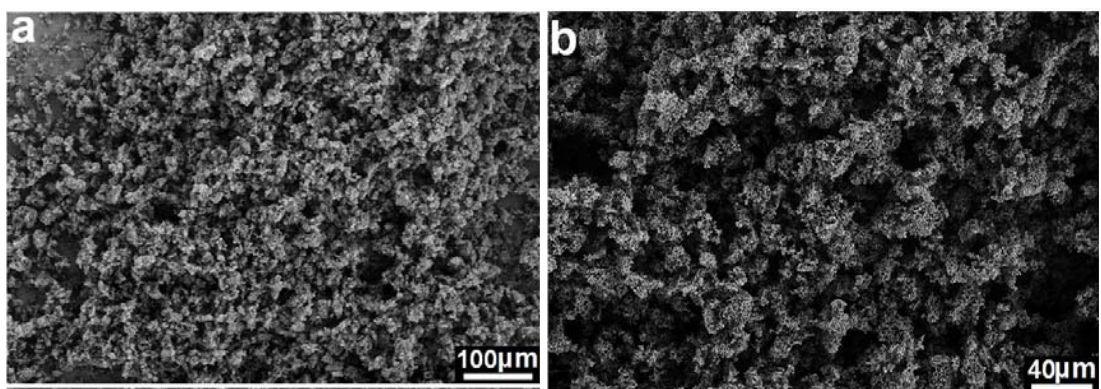

**Fig. S4.** SEM images of PNCF in low magnifications.

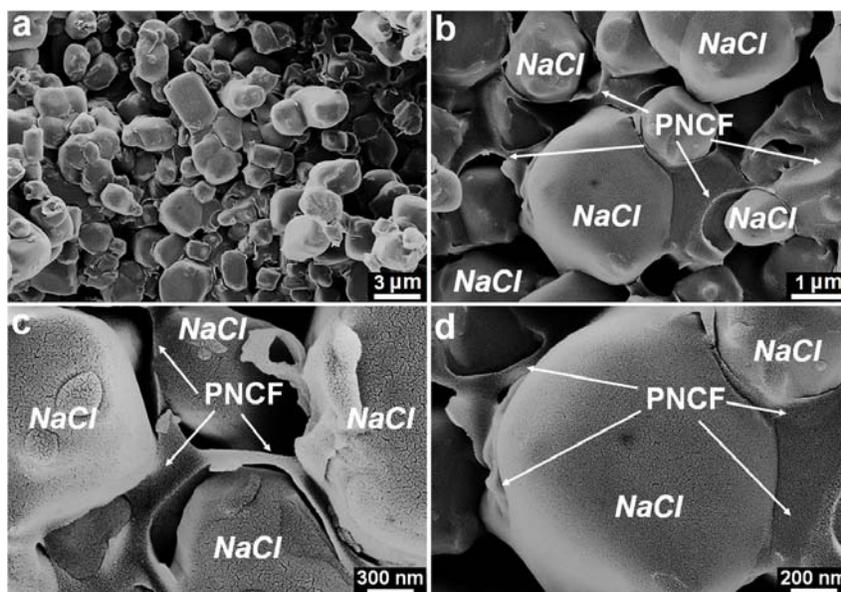

**Fig. S5.** SEM images in different magnifications of NaCl/nitrogen-doped carbon composite obtained from the carbonization of NaCl/PILMI mixture at 750 °C before water washing treatment to remove the NaCl template.

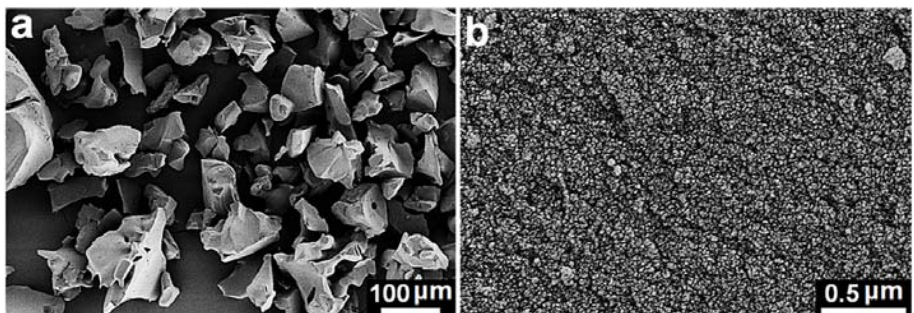

**Fig. S6.** SEM images in low (a) and high (b) magnifications of the C-PILMI prepared from the carbonization of merely PILMI in the absence of NaCl template.

No macrosopic pores are observed in the C-PILMI, and it contains many large macroparticles with a size of 50–200 μm. Each large macroparticle consists of numerous nanoparticles with a size of 30–50 nm.

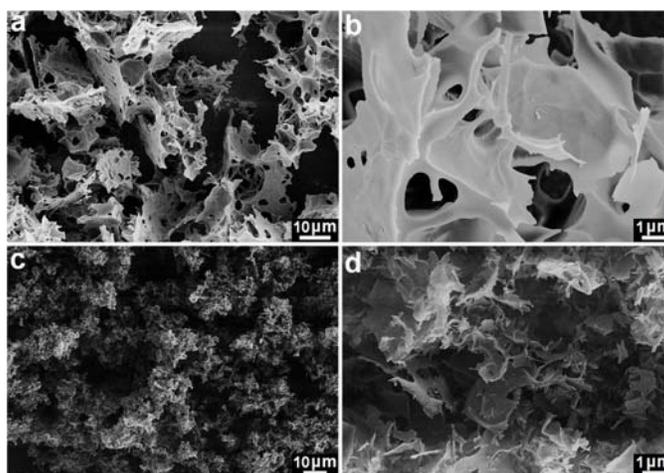

**Fig. S7.** SEM images of carbon products at the NaCl/PILMI mass ratio of 2 (a and b, irregular carbon with few macropores) and 50 (c and d, randomly distributed carbon nanosheets).

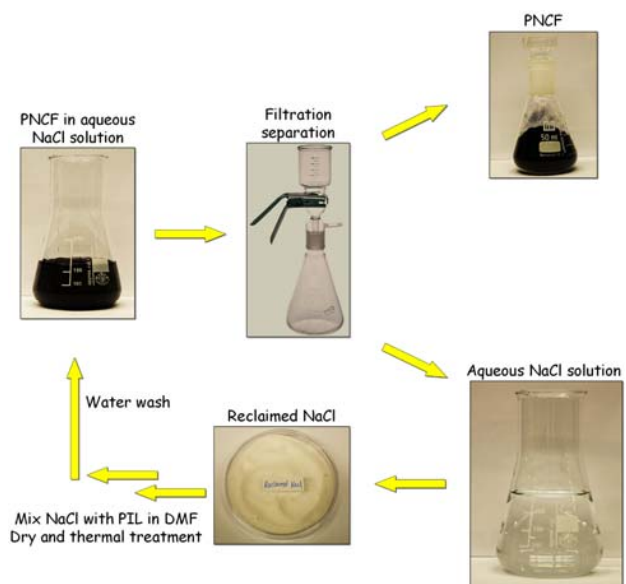

**Fig. S8.** The reclaim process of NaCl in the "solid salt template" strategy.

In this process, about 95 wt % of NaCl can be reclaimed. We also find that PNCF prepared by using the reclaimed NaCl shows similar morphology and microstructure with that prepared by using fresh NaCl. In this regards, our "solid salt templating" approach by using NaCl is superior to the traditional template methods such as by using $SiO_2$, $CaCO_3$ and nickel foam, which usually are hard to be reused.

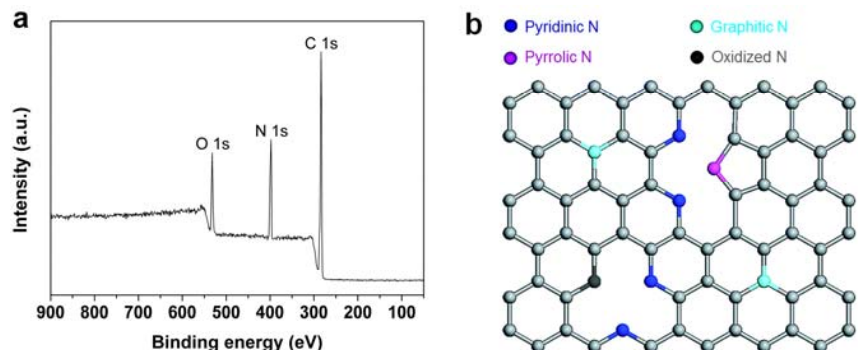

**Fig. S9.** (a) XPS curve of PNCF and (b) scheme of N element of different chemcial states on the surface of PNCF.

High-resolution N 1s XPS curve is deconvoluted into four components: pyridinic nitrogen (398.3 eV, 43.9%), pyrrolic nitrogen (399.8 eV, 12.4%), graphitic nitrogen (400.9 eV, 26.4%) and oxidized nitrogen (402.8 eV, 7.3%), according to the previous work [36, 37].

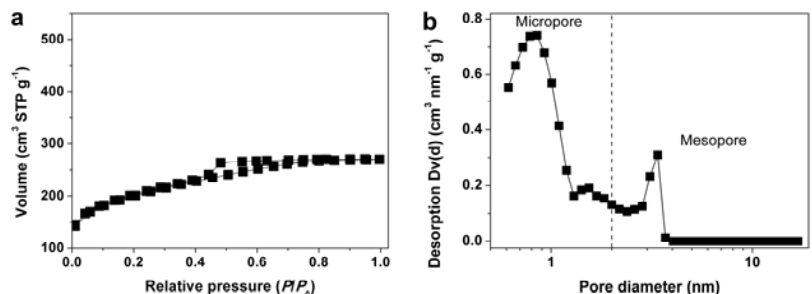

**Fig. S10.** (a) Nitrogen adsorption/desorption isotherms at 77 K and (b) the corresponding pore size distribution plot of C-PILMI prepared from the carbonization of merely PILMI at 750 °C in the absence of NaCl template.

C-PILMI presents a combined type I/IV physisorption isotherm. A high adsorption capacity at the low relative pressure $P/P_0 < 0.1$ is observed, which reveals the significant presence of micropores in the C-PILMI. The detectable type-H4 hysteresis loop at the relative pressure $P/P_0$ ranging from 0.4 to 0.8 corresponds to the filling and emptying of a small fraction of mesopores by capillary condensation. The "self-template" feature of PILMI, that is to say, the formation mechanism of micropores and small mesopores, is believed to be similar to that reported by Dai *et al* [38]. They found that cyano group in the chemical structure of the ionic liquids underwent trimerization reaction at medium temperatures (350–450 °C), firstly producing a solid triazine-based polymeric network. The large-sized anions (here $Tf_2N$) were trapped and aggregated in the matrix, and left behind the nanopores *via* the subsequent volatilization of this entity.

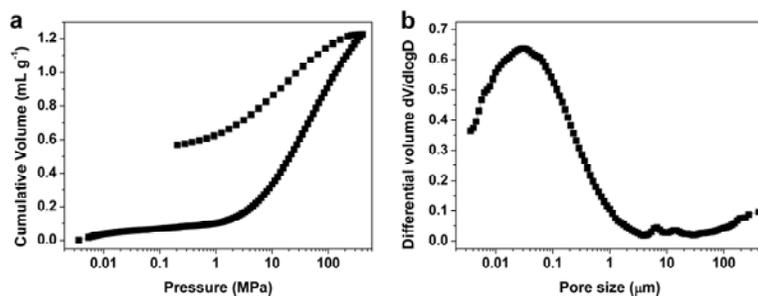

**Fig. S11.** (a) Cumulative mercury volume as a function of the pressure, and (b) the corresponding differential pore volume as a function of the pore size.

To clearly show the large mesopores and macropores of PNCF, the mercury porosimetry experiment is conducted. It is shown that PNCF displays a lot of large mesopores and macropores in the range of 10 nm–3 μm. It should be noted that in order to facilitate the measurement, the PNCF powder is pressed into a table before the measurement.

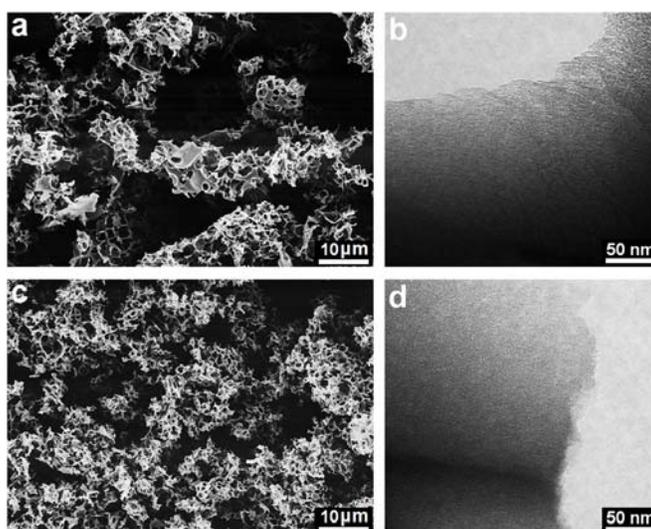

**Fig. S12.** SEM and TEM images of hierarchically porous nitrogen-doped carbon foams produced by employing (a and b) KCl (melting point = 770 ºC) or (c and d) $Na_2SO_4$ (melting point = 884 ºC) as the solid salt template and PILMI as the precursor at 725 and 750 ºC, respectively.

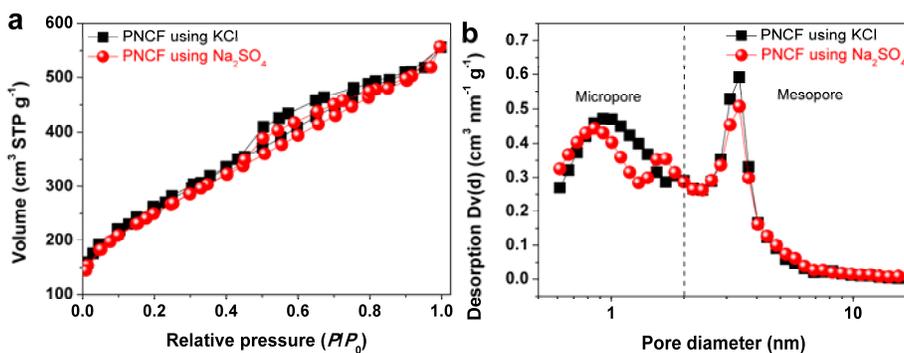

**Fig. S13.** (a) Nitrogen adsorption/desorption isotherms at 77 K and (b) the corresponding pore size distribution plots of the hierarchically porous nitrogen-doped carbon foams produced by using KCl or $Na_2SO_4$ as the solid salt template and PILMI as the precursor.

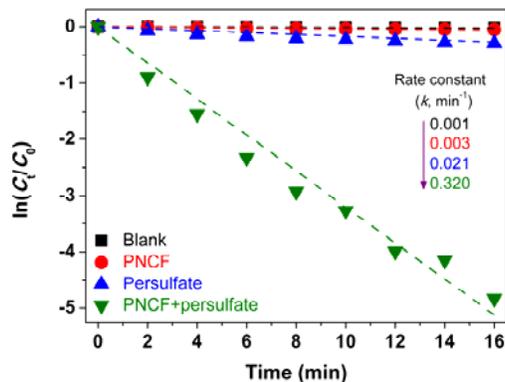

**Fig. S14.** Linear fitting between ln($C_t/C_0$) and irradiation time in different systems.

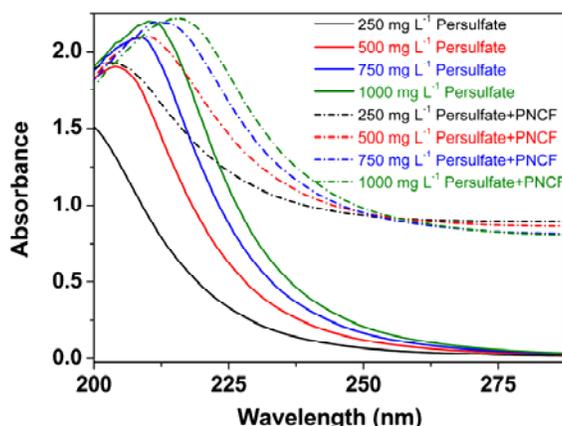

**Fig. S15.** UV-Vis absorbance of different persulfate solutions before and after the PNCF addition (60 mg L$^{-1}$). A red shift of UV-Vis absorbance of persulfate solution is observed after adding PNCF, suggesting the electrostatic and $\pi$-$\pi$ interactions of PNCF and persulfate.

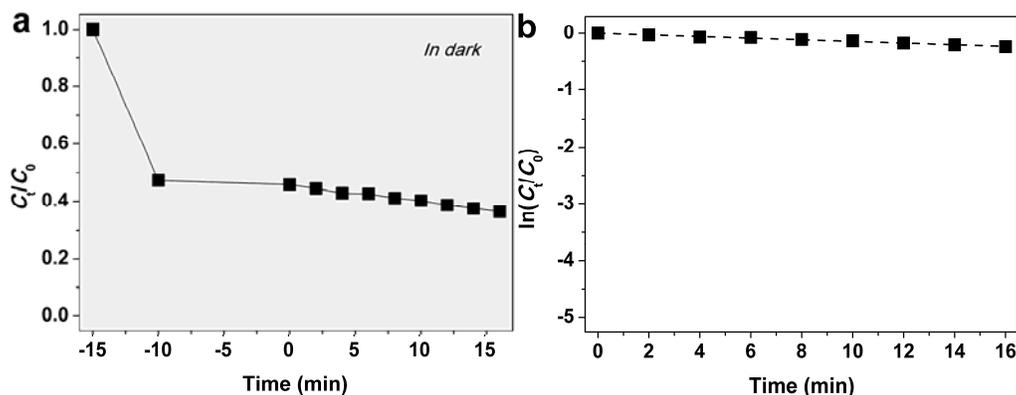

**Fig. S16.** Catalytic degradation of RB in the PNCF + persulfate system **in dark**, and (b) its linear fittings between ln($C_t/C_0$) and reaction time. The apparent rate constant ($k$, min$^{-1}$) of RB degradation in the PNCF + persulfate system in dark is 0.014 min$^{-1}$, which is slower than that in the PNCF + persulfate system under visible light irradiation (0.320 min$^{-1}$) and that in the persulfate system under visible light irradiation (0.021 min$^{-1}$). These results indicate that NPCF itself can to an extent enhance the activation of persulfate, presumably due to the $sp^2$ carbon in graphitic layers, the zigzag edges with unconfined $\pi$ electrons and the doping nitrogen atom, which act as active sites. Besides, the activation efficiency of persulfate by NPCF in dark is lower than that by through the excited RB molecule under visible light irradiation. That is to say, the excited RB molecule contributes more than NPCF to the activation of persulfate.

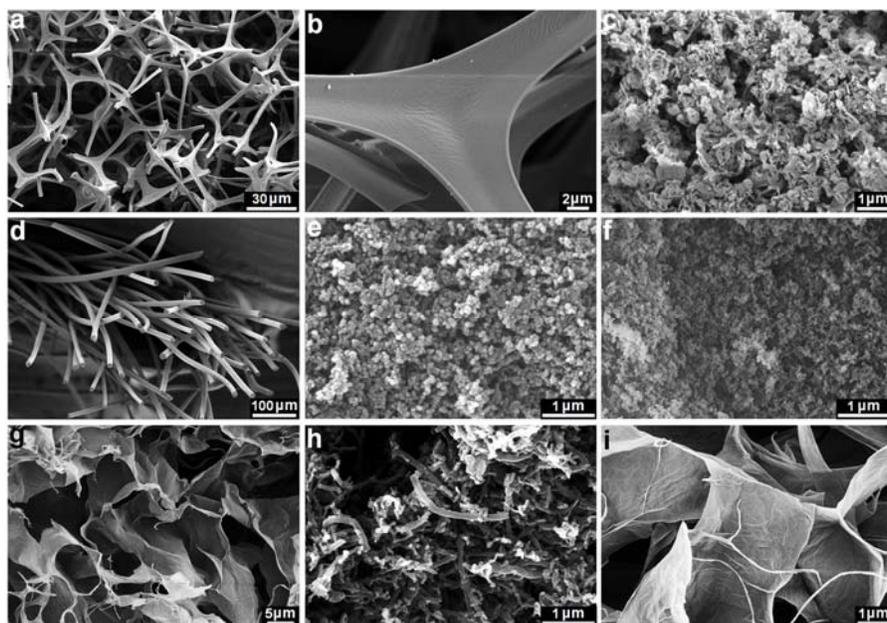

**Fig. S17.** SEM images of some representative carbonaceous materials, including (a and b) nitrogen-doped carbon fiber (NCF), (c) carbon nitride ($C_3N_4$), (d) porous carbon fiber (PCF), (e) acetylene carbon (AC), (f) fullerene (C60), (g) graphene (G), (h) CNT and (i) graphene oxide (GO).

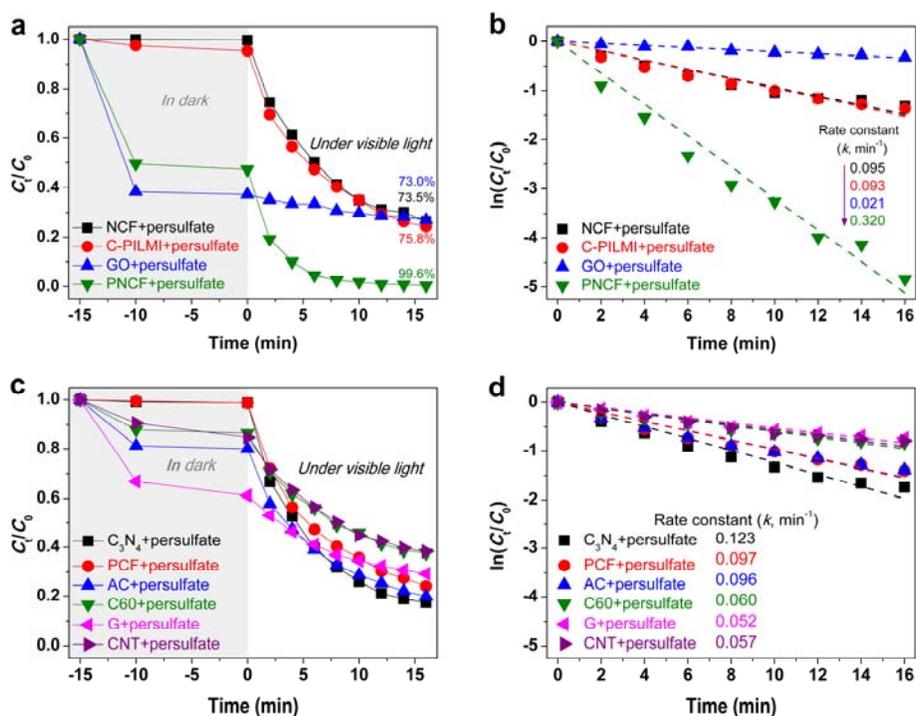

**Fig. S18.** (a) Catalytic degradation of RB (20 mg mL$^{-1}$) under visible light irradiation over different catalysts such as NCF, C-PILMI, GO and PNCF, and (b) their linear fittings between $\ln(C_t/C_0)$ and irradiation time. (c) Catalytic degradation of RB under visible light irradiation over different catalysts including $C_3N_4$, PCF, AC, C60, G and CNT, and (d) their linear fittings between $\ln(C_t/C_0)$ and irradiation time. The apparent rate constant ($k$, min$^{-1}$) of RB degradation over different catalysts is listed in (b) and (d).

In the system of the persulfate + carbon material (including GO, PCF, AC, C60, G and CNT) containing no nitrogen dopants, these carbon materials bear the $sp^2$ carbon in graphitic layers and the zigzag edges with unconfined $\pi$ electrons, which act as active sites and contribute to the activation of persulfate. In the system of the persulfate + carbon-based material (including PNCF, NCF, C-PILMI and $C_3N_4$) containing nitrogen dopants, these nitrogen dopants bear higher electron density than carbon atoms and provide extra active sites for the activation of persulfate.

For NCF and $C_3N_4$, the low $S_{BET}$ lows down their overall degradation efficiency of RB (that is, they could not effectively pre-concentrate the persulfate and RB molecules before the activation of persulfate and the degradation of RB), compared to PNCF. For C-PILMI, despite its high $S_{BET}$ of 708 m$^2$ g$^{-1}$, the absence of rich mesopores and macropores leads to the low capacity and slow diffusion of persulfate and/or RB molecules, and thus results in the inferior degradation efficiency of RB, compared to PNCF.

It should be noted that although NCF bears a greatly lower $S_{BET}$ of 92 m$^2$ g$^{-1}$ than AC (240 m$^2$ g$^{-1}$) or PCF (801 m$^2$ g$^{-1}$), the apparent rate constant ($k$) of RB for NCF (0.95 min$^{-1}$) is comparable to that for AC (0.96 min$^{-1}$) or PCF (0.97 min$^{-1}$), which confirms the important role of nitrogen dopant in the activation of persulfate.

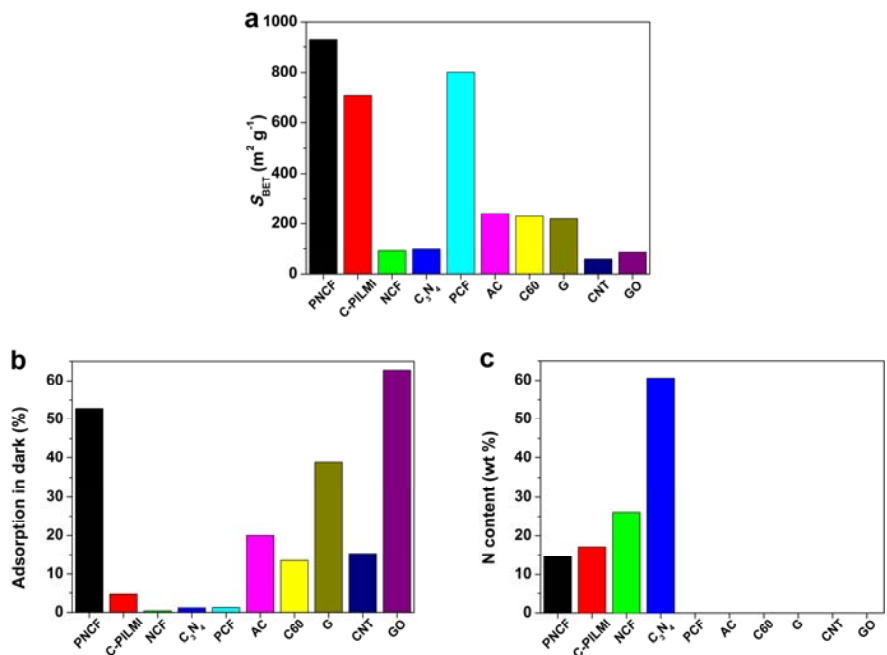

**Fig. S19.** Comparisons of (a) $S_{BET}$, (b) the RB adsorption efficiency in dark, and (c) N content of different catalysts, including PNCF, C-PILMI, NCF, $C_3N_4$, PCF, AC, C60, G, CNT and GO.

**References to Supplementary Material**
[1] X. Li, Z. Ao, J. Liu, H. Sun, A.I. Rykov, J. Wang, Topotactic transformation of metal-organic frameworks to graphene-encapsulated transition-metal nitrides as efficient Fenton-like catalysts, ACS Nano 10 (2016) 11532-11540.
[2] H. Dong, Z. Qiang, J. Hu, C. Sans, Accelerated degradation of iopamidol in iron activated persulfate systems: Roles of complexing agents, Chem. Eng. J. 316 (2017) 288-295.
[3] Y. Yao, Y. Cai, G. Wu, F. Wei, X. Li, H. Chen, S. Wang, Sulfate radicals induced from peroxymonosulfate by cobalt manganese oxides ($Co_xMn_{3-x}O_4$) for fenton-like reaction in water, J. Hazard. Mater. 296 (2015) 128-137.